\documentclass[floatfix,twocolumn,aps,prb,showpacs,amsmath,amssymb]{revtex4-1}
\usepackage[latin1]{inputenc}
\usepackage[dvips]{graphicx}

\usepackage{dcolumn}
\usepackage{bm}
\usepackage{dsfont}
\usepackage{color}
\usepackage{url}
\usepackage[colorlinks=true ,citecolor=blue]{hyperref}
\usepackage{amsmath,amssymb,esint}
\usepackage[table]{xcolor}

\definecolor{Nathanblue}{rgb}{0.,0.24,0.51}

\newcommand{\be}{\begin{equation}}
\newcommand{\ee}{\end{equation}}
\newcommand{\bq}{\begin{eqnarray}}
\newcommand{\eq}{\end{eqnarray}}

\newcommand{\tr}{\mathrm{tr}}

\begin{document}

\title{Non-Abelian tensor Berry connections in multi-band topological systems}

\author{Giandomenico Palumbo}
\affiliation{School of Theoretical Physics, Dublin Institute for Advanced Studies, 10 Burlington Road,
	Dublin 4, Ireland}

\date{\today}

\begin{abstract}

\noindent Here, we introduce and apply non-Abelian tensor Berry connections to topological phases in multi-band systems. These gauge connections behave as non-Abelian antisymmetric tensor gauge fields in momentum space and naturally generalize Abelian tensor Berry connections and ordinary non-Abelian (vector) Berry connections.
We build these novel gauge fields from momentum-space Higgs fields, which emerge from the degenerate band structure of multi-band models. Firstly, we show that the conventional topological invariants of two-dimensional topological insulators and three-dimensional Dirac semimetals can be derived from the winding number associated to the Higgs field. Secondly, through the non-Abelian tensor Berry connections we construct higher-dimensional Berry-Zak phases and show their role in the topological characterization of several gapped and gapless systems, ranging from two-dimensional Euler insulators to four-dimensional Dirac semimetals. Importantly, through our new theoretical formalism, we identify and characterize a novel class of models that support space-time inversion and chiral symmetries. 
Our work provides an unifying framework for different multi-band topological systems and sheds new light on the emergence of non-Abelian gauge fields in condensed matter physics, with direct implications on the search for novel topological phases in solid-state and synthetic systems. 

\end{abstract}

\maketitle

\noindent {\bf Introduction:} Non-Abelian Berry connections \cite{Berry,Zak,Wilczek} play a central role in multi-band systems with degenerate spectra. These connections behave like non-Abelian vector gauge fields in momentum/parameter space and have been applied in different research areas ranging from quantum computation \cite{Pachos,Pachos2} to topological phases of matter \cite{Niu3,Bernevig4}.
The topology of several multi-band systems is naturally encoded in non-Abelian Berry connections which are responsible for the quantum spin Hall effect in 2D \cite{Dai}, the electric polarization in 3D \cite{Qi, Moore}  and for the second Chern number in 4D \cite{Qi,Kolo,Rastelli}.
Non-Abelian Berry connections give rise to Wilson loops, which are a powerful tool of investigation in topological matter \cite{Dai,Bernevig3,Benalcazar, Bradlyn2,Slager,Goldman,Spielman}. 
Moreover, topological Bloch oscillations in topological crystalline insulators \cite{Holler} and higher-order topological insulators \cite{Palumbo5} are naturally related to the existence of non-Abelian Berry connections, which influence the dynamics of wave-packets \cite{Niu2,Shindou}.
Importantly, a generalization of Abelian Berry connections have been recently proposed \cite{Palumbo3,Palumbo2}, where the new connections behave like Abelian antisymmetric tensor (Kalb-Ramond \cite{Kalb}) gauge fields in momentum/parameter space.
These tensor Berry connections have been employed to characterise the topology of 3D chiral topological insulators \cite{Palumbo2} and 4D topological semimetals \cite{Palumbo3,Palumbo4,Zhu} where the Dixmier-Douady (DD) invariant replaces the Chern number. The theoretical developments recently let to the experimental measurement of the DD invariant in 4D synthetic systems \cite{Yu,Cappellaro}.

The main goal of our paper is to unveil the existence of novel non-Abelian Berry connections, coined \emph{non-Abelian tensor Berry connections}, which behave like non-Abelian antisymmetric gauge fields in the momentum space of multi-band systems with degenerate spectra. These types of tensor fields naturally appear in high-energy physics \cite{Freedman,Smolin} where they are defined in real spacetime, while in mathematical literature they are known as non-Abelian gerbe connections \cite{Murray, Murray2,Thiang}.
Here, we build these new connections by combining the conventional non-Abelian Berry curvature together with non-Abelian scalar fields that behave as momentum-space Higgs fields. Within the framework of topological phases of matter, 
considering several gapless and gapped systems in different dimensions, we will show that these Higgs fields and non-Abelian tensor Berry connections emerge from band structures, hence allowing us to derive their topological bulk invariants.\\
Finally, we will show the existence of a novel class of topological phases characterized by topological invariants associated to non-Abelian real bundle gerbes. These systems defined for gapped $(2n+1)$-D and gapless $(2n+2)$-D models are characterised by space-time inversion and chiral symmetries, which give rise to degenerate spectra with real Bloch wavefunctions. We will provide two explicit models in 3D and 4D, respectively, where their topological invariants are directly related to $SO(4)$ tensor Berry connections.
 Importantly, through these new tensor connections, novel higher-dimensional models (with an eventual higher number of bands) in this class can be easily identified.\\
Our work provides an unifying theoretical framework for different multi-band topological systems and our results shed light on the existence of novel gauge structures and geometric phases in quantum matter, with important implications on the search for novel topological phases in solid-state and synthetic setups.

\noindent {\bf Momentum-space Higgs field: } We start by generalizing the construction of momentum-space Higgs fields in multi-band models. These non-Abelian scalars will play a central role in the definition of non-Abelian tensor Berry connections as will show in the next section. We note that non-Abelian complex scalar fields in parameter space have been previously considered in the context of fictitious 't Hooft-Polyakov monopoles \cite{Tong,Niu,Hashimoto,Ohya}. However, their formulation in multi-band topological models and their role in topological phases of matter have never been analyzed in detail. The main goal of this section is thus to fill this gap. Given a degenerate Bloch state $|u({\bf{k}})\rangle=(|u_1({\bf{k}})\rangle, |u_2({\bf{k}})\rangle,...,|u_N({\bf{k}})\rangle)^T$ associated to $N$ degenerate bands with energy $E(\bf{k})$, with $\bf{k}$ referring to the momenta of a generic lattice system, we define the matrix components $\tilde{\Phi}_{ab}$ of a non-Abelian scalar field $\tilde{\Phi}$ as follows
\begin{eqnarray}\label{}
\tilde{\Phi}_{ab} = \langle u_a|G|u_b\rangle,
\end{eqnarray}
where $\{a,b\}=1,...,N$. Here, $\tilde{\Phi}= \tilde{\phi}_i T^i$, where $\tilde{\phi}_i$ are the vector components of $\tilde{\Phi}$ and $T^i$ are the generators of a $U(N)$, $SU(N)$ or $SO(N)$ Lie algebra. 
Importantly, the matrix $G$ is related to certain symmetries of the given lattice model. In particular, all the systems discussed in this paper share (besides time-reversal symmetry) the 2D inversion symmetry, namely $\mathcal{I}_{2D} H(k_1, k_2,k_3,..., k_n) \mathcal{I}_{2D}^{-1} = H (-k_1, -k_2,k_3,..., k_n)$,
where, for simplicity, we identify $k_1$ and $k_2$ with $k_x$ and $k_y$, respectively. In this way, we will take 
$G\equiv \mathcal{I}_{2D}$ in the rest of the paper (clearly, the representation of $\mathcal{I}_{2D}$ is model dependent).
Moreover, we have that $\tilde{\Phi}\cdot \tilde{\Phi}=\Lambda \mathbb{I}$, where $\mathbb{I}$ is the $N \times N$ identity matrix, the dot symbol $\cdot$ represents the the matrix product and $\Lambda$ is a momentum-dependent normalization factor.
We then define the corresponding normalized quantity given by
\begin{eqnarray}\label{}
\Phi = \frac{\tilde{\Phi}}{\sqrt{\Lambda}},  \hspace{0.5cm} \Phi \cdot \Phi = \mathbb{I}.
\end{eqnarray}
Here, $\Phi$ represents the \emph{momentum-space Higgs field} that we will employ in the rest of the paper. However, before introducing the non-Abelian tensor Berry connections, we first propose to show that the topology of the Higgs field has relevant implications in topological phases.
In quantum field theory, the topology of the Higgs field in three-dimensional space is given by the winding number \cite{Freund,Tchrakian,Kibble}
\begin{eqnarray}\label{2DwindingS}
w(\mathbb{S}^2) = \frac{1}{16 \pi i }\int_{\mathbb{S}^2}dS_i\,\epsilon^{ijk}\, {\rm \tr}\left(\Phi \cdot \partial_j \Phi \cdot \partial_k \Phi\right),
\end{eqnarray}
which represents the magnetic charge on the two-dimensional sphere $\mathbb{S}^2$ related to the residual U(1) gauge invariance of 't Hooft-Polyakov monopoles. In a similar way, we can construct the winding number on the two-dimensional torus $\mathbb{T}^2$ (i.e. the first Brillouin zone of a lattice system).
\begin{figure}[]
	\center
	\includegraphics[width=1.0\columnwidth]{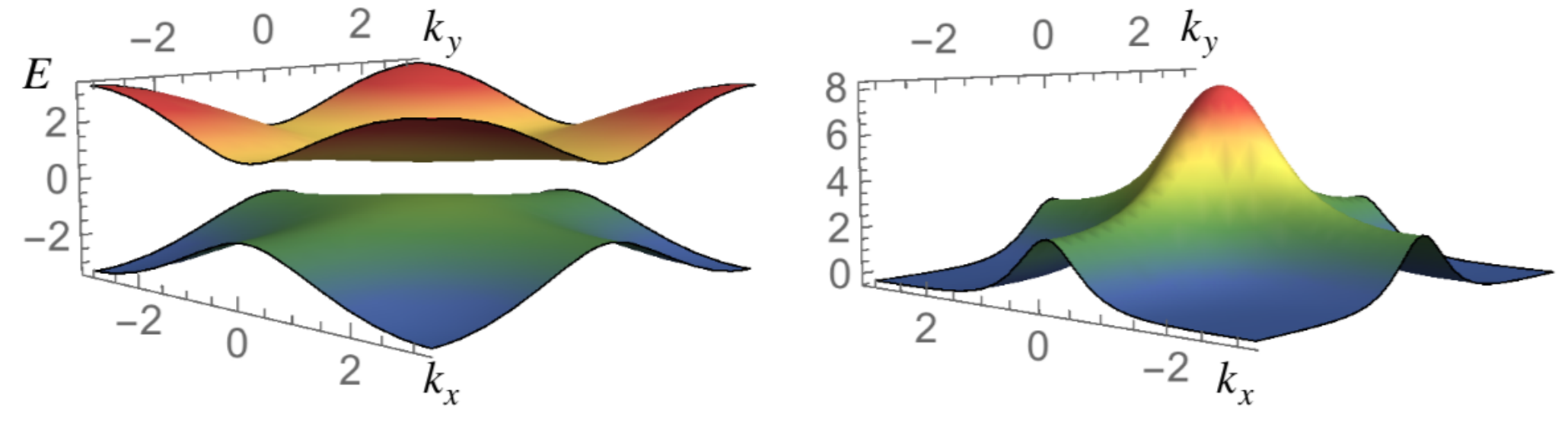}
	\caption{(Left) Band structure of the topological phase in the BHZ model at $m=1.3$. (Right) Plot of the integrand of the winding number $w(\mathbb{T}^2)$ for the BHZ model at $m=1.3$. The maximum of the function is at the $\Gamma$ point.
	}
	\label{fig1}
\end{figure}
To give a first concrete application of the winding number associated to the Higgs field, we consider the Bernevig-Hughes-Zhang (BHZ) model for 2D quantum spin Hall (QSH) insulators \cite{Bernevig}. The corresponding momentum space Hamiltonian is given by
\begin{align}\label{BHZ}
H_{BHZ}= \gamma_1 \sin k_x + \gamma_2 \sin k_y+\gamma_3 (m-\cos k_x-\cos k_y),
\end{align}
where $\gamma_1=-\sigma_y \otimes \sigma_z$, $\gamma_2=-\sigma_y \otimes \sigma_x$ and $\gamma_3=\sigma_x \otimes \sigma_0$ and $\sigma_i$ the Pauli matrices (here, we have adopted a chiral-invariant basis). Its degenerate spectrum is given by
\begin{align}\label{}
E_{\pm}= \pm \sqrt{2+m^2-2m(\cos k_x+\cos k_y)+2\cos k_x \cos k_y},
\end{align}
and at half filling, the time-reversal-invariant topological phase holds for $|m|<2$ ($m\neq0$) for which the $\mathbb{Z}_2$ invariant is 1. We can now build the Higgs field for this model by taking $G=\sigma_0 \otimes \sigma_y$ (which is related to $\mathcal{I}_{2D}$) and $|u\rangle$ the doubly degenerate Bloch state associated to the lower occupied bands
\begin{eqnarray}\label{}
\Phi=\frac{\sigma_i f^i}{|E|},
\end{eqnarray}
with $f^x=\sin k_x$, $f^y=m-\cos k_x-\cos k_y$ and $f^z=\sin k_y$.
By applying the formula for the winding number in Eq.(\ref{2DwindingS}) on $\mathbb{T}^2$, we obtain
\begin{align}\label{}
w(\mathbb{T}^2) = \int_{\mathbb{T}^2}d^2k\, \frac{\cos k_x + \cos k_y-m \cos k_x \cos k_y}{4 \pi |E|^3}= \nonumber \\
={\rm sign}(m), \hspace{0.3cm} |m|<2, \hspace{1.5cm}
\end{align}
with $w(\mathbb{T}^2)=0$ for $|m|>2$. Thus we recognize the absolute value of the winding number $|w(\mathbb{T}^2)|$ as the $\mathbb{Z}_2$ invariant in 2D QSH insulators. Notice that the integrand of the above winding number has a maximum at the $\Gamma$ point, i.e. in correspondence with the gap closing point at the topological phase transition ($|m|=2$) of the BHZ model as shown in Fig.(\ref{fig1}).\\
The winding number on the sphere $w(\mathbb{S}^2)$ has also relevant applications in gapless topological phases. In this case, we consider the linearized momentum-space Hamiltonian of a 3D Dirac semimetal \cite{Kane} given by
\begin{eqnarray}\label{3DDirac}
H_{3D}= \gamma_1  k_x + \gamma_2  k_y+\gamma_3  k_z,
\end{eqnarray}
with $\gamma_1$, $\gamma_2$ and $\gamma_3$ the same Dirac matrices defined in Eq.(\ref{BHZ}). The corresponding Higgs field $\Phi$, built from the same $G$-matrix as for the previous model, is given by
\begin{eqnarray}\label{}
\Phi=- \frac{k_x \sigma_x+k_y \sigma_z+k_z \sigma_y}{|k|},
\end{eqnarray}
such that
\begin{eqnarray}\label{}
w(\mathbb{S}^2) = \frac{1}{2 \pi }\int_{\mathbb{S}^2}dS_i\,\frac{k_i}{|k|^3}=2.
\end{eqnarray}
This shows that the 3D Dirac point behaves like a momentum-space monopole \cite{Goswami,Wieder4} and that $|w(\mathbb{S}^2)|$ represents its topological charge. We point out that this result can be naturally extended to $N$-degenerate Dirac cones \cite{Bradlyn} where $|w(\mathbb{S}^2)| =N$, such as in double Dirac semimetals with $N=4$ \cite{Wieder2,Bouhon}.\\
 
\noindent {\bf Non-Abelian tensor Berry connections: } Through the momentum-space Higgs field $\Phi$ we can now build novel gauge connections that we coin \emph{non-Abelian tensor Berry connections} ${\bf{B}}_{ij}$. They are defined as follows
\begin{eqnarray}\label{TensorBerry}
{\bf{B}}_{ij}= \Phi\, \cdot {\bf{F}}_{ij},
\end{eqnarray}
where ${\bf{F}}_{ij}$ is the non-Abelian Berry curvature \cite{Niu2,Shindou}
\begin{eqnarray}\label{}
{\bf{F}}_{ij}= \partial_{i} {\bf{ A}}_j - \partial_{j} {\bf {A}}_i -i [{\bf{A}}_i, {\bf{A}}_j],
\end{eqnarray}
with ${\bf {A}}_i$ the non-Abelian Berry connection and $\partial_{i}\equiv \partial_{k_i}$.
Under gauge transformations $|u\rangle\rightarrow U |u\rangle$ we have that
\begin{eqnarray}\label{}
{\bf{F}}_{ij}\rightarrow U \cdot {\bf{F}}_{ij} \cdot U^{-1}, \hspace{0.3cm} \Phi\rightarrow U \cdot \Phi \cdot U^{-1},
\end{eqnarray}
where $U$ is a  Lie-algebra-valued
matrix, such that ${\bf{B}}_{ij}$ also transforms in a gauge-covariant way; we note that the trace over the degenerate-band indices of all these quantities is gauge invariant. The tensor gauge field ${\bf{B}}_{ij}$ behaves as a non-Abelian Kalb-Ramond field \cite{Freedman,Smolin,Spallucci} in momentum space and its curvature tensor is given by
\begin{eqnarray}\label{}
{\bf H}_{ijk}={\bf{D}}_i {\bf{B}}_{jk}+{\bf{D}}_j {\bf{B}}_{ki}+{\bf{D}}_k {\bf{B}}_{ij},
\end{eqnarray}
such that
\begin{eqnarray}\label{}
{\bf{H}}_{ijk}\rightarrow U\cdot {\bf{H}}_{ijk}\cdot U^{-1},
\end{eqnarray}
with ${\bf{D}}_j f=i\, \partial_j f- [{\bf{ A}}_j, f]$ the covariant derivative (here, $f$ is a generic Lie-algebra valued function).
Furthermore, there exist natural generalizations of ${\bf{B}}_{ij}$, named C-fields, which are three-form gauge fields \cite{Johnson}. Thus,
a \emph{non-Abelian higher-tensor Berry connection} can be built from ${\bf H}_{ijk}$ as follows
\begin{eqnarray}\label{}
{\bf C}_{ijk}=\Phi\, \cdot {\bf H}_{ijk},
\end{eqnarray}
which gives rise to its own higher curvature tensor. Similarly to ${\bf{B}}_{ij}$, ${\bf C}_{ijk}$ also transforms in a gauge-covariant way and its trace is gauge invariant.
Antisymmetric tensor fields as ${\bf{B}}_{ij}$ and ${\bf C}_{ijk}$ are known in differential geometry and topology as gerbe connections, which are related to bundle gerbes \cite{Murray} and higher bundle gerbes \cite{Murray2,Johnson}, respectively. These structures naturally generalize fiber bundles \cite{Nakahara}. Moreover, similarly to Wilson loops, it is possible to build from these gauge connections non-local gauge operators, named Wilson surfaces \cite{Waldorf,Picken} and Wilson volumes. 
For our purpose, the simplest gauge invariant quantities associated to ${\bf{B}}_{ij}$ and ${\bf C}_{ijk}$ are respectively given by
\begin{eqnarray}\label{}
\Upsilon_{\bf{B}}(\mathbb{M}^2) = \frac{1}{2 \pi }\int_{\mathbb{M}^2} d^2k \, {\rm \tr}\, {\bf{B}}_{xy}, \hspace{0.23cm}\nonumber \\
\Upsilon_{\bf{C}}(\mathbb{M}^3) = \frac{1}{(2 \pi)^2 }\int_{\mathbb{M}^3} d^3k \, {\rm \tr}\, {\bf{C}}_{xyz},
\end{eqnarray}
where $\mathbb{M}^2$ and $\mathbb{M}^3$ are 2D and 3D compact manifolds, respectively. These expressions naturally generalize non-Abelian Berry-Zak phases \cite{Zak} in higher dimensions.\\
We now revisit some known topological phases and show that their topological invariants can be formulated in terms of the higher-dimensional non-Abelian Berry-Zak phases. 
2D Euler insulators are the prototypical example of topological phases characterized by space-time inversion $\mathcal{IT}$ ($\mathcal{C}_2 \times \mathcal{T}$) with time-reversal symmetry $\mathcal{T}^2=1$ and an Euler number induced by a $SO(2)$ Berry connection \cite{Ahn2,Unal,Zhao4,Bouhon2,Slager2,Zhao5,Ezawa}. A simple model with four bands is given by the BHZ Hamiltonian in Eq.(\ref{BHZ}), but now with the Dirac matrices in a real representation of the Clifford algebra: $\gamma_1=\sigma_z \otimes \sigma_0$, $\gamma_2=\sigma_y \otimes \sigma_y$ and $\gamma_3=\sigma_x \otimes \sigma_0$. To construct the Higgs field, we consider $G=\sigma_x \otimes \sigma_y$ such that $\Phi=-\sigma_y$ for the two degenerate lower bands and obtain
\begin{eqnarray}\label{}
\Upsilon_{\bf{B}}(\mathbb{T}^2) = \frac{1}{2 \pi }\int_{\mathbb{T}^2} d^2k \, {\rm \tr}\, {\bf{B}}_{xy} = 2\,{\rm sign}(m), \hspace{0.1cm} |m|<2, \hspace{0.3cm}
\end{eqnarray}
with $\Upsilon_{\bf{B}}(\mathbb{T}^2)=0$ for $|m|>2$. Notice that $\Upsilon_{\bf{B}}$ coincides with topological Euler invariant $e_1$, while ${\bf{B}}_{xy}$ is mathematically equivalent to a gauge connection for a real bundle gerbe \cite{Mathai,Szabo}. As we will see more in detail in the next section, real bundle gerbes will play a central role in the characterization of higher-dimensional topological phases with real Bloch wavefunctions.
To reveal the role of ${\bf C}$ connections, we consider the linearized Hamiltonian of a 4D Dirac semimetal described by the following momentum-space Hamiltonian
\begin{eqnarray}\label{4DDirac}
H_{4D}= \gamma_1 k_x + \gamma_2 k_y+\gamma_3 k_z+\gamma_4 k_w,
\end{eqnarray}
where we have employed the same Dirac matrices as for the 3D Dirac semimetal in the previous section together with $\gamma_4=-\sigma_y \otimes \sigma_y$. We derive the corresponding Higgs field $\Phi$ with $G=\sigma_0 \otimes \sigma_y$ (which is associated to the 2D inversion for $k_x$ and $k_y$) and obtain the following topological invariant
\begin{eqnarray}\label{}
\Upsilon_{\bf{C}}(\mathbb{S}^3) = \frac{1}{(2 \pi)^2 }\int_{\mathbb{S}^3} dk^i \wedge dk^j \wedge dk^k \, {\rm \tr}\, {\bf{C}}_{ijk} = 2,
\end{eqnarray}
which we thus identify as the $\mathbb{Z}_2$ invariant of the 4D Dirac point. So far, we have discussed topological phases that were already studied in the literature. In the next section, we will employ the non-Abelian tensor Berry connections to unveil the existence of new topological states in 3D and 4D.\\

\begin{figure}[]
	\center
	\includegraphics[width=1\columnwidth]{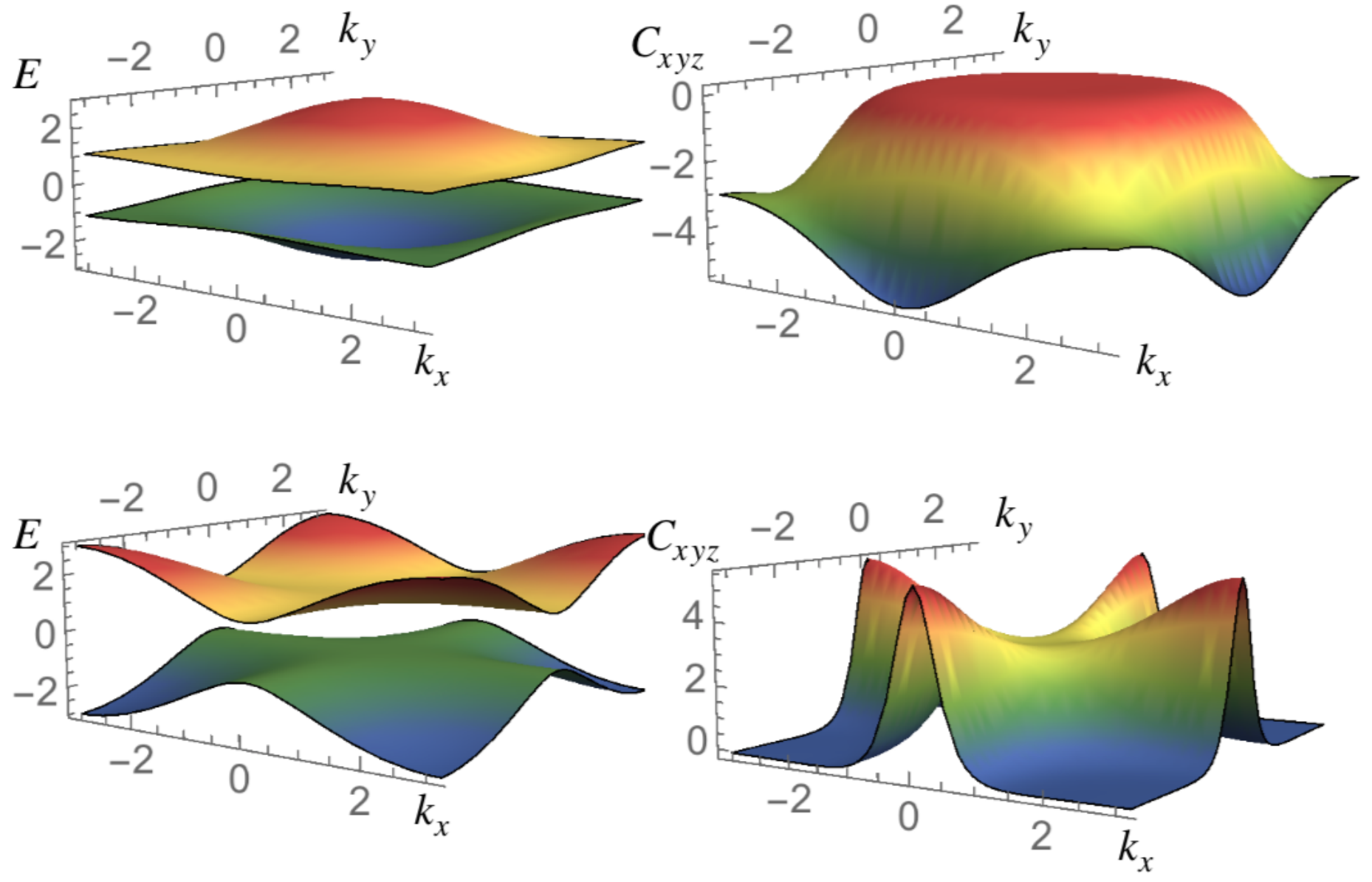}
	\caption{(Left-Up) Band structure of the 3D gapped topological phase with $\mathcal{IT}$ and $\mathcal{S}$ symmetries at $m=0.1$ and $k_z=0$. (Right-Up) Plot of ${\bf{C}}_{xyz}$ associated to $\Upsilon_{\bf{C}}(\mathbb{T}^3)=-8$ at $m=0.1$. The maximum of the function is at $\Gamma$ point.
		(Left-Down) Band structure of the same model at $m=1.9$ and and $k_z=0$. (Right-Down) Plot of ${\bf{C}}_{xyz}$ associated to $\Upsilon_{\bf{C}}(\mathbb{T}^3)=4$ at $m=1.9$. The maximum of the function is at $X$ points.
	}
	\label{fig2}
\end{figure}

\noindent {\bf Topological phases with space-time inversion and chiral symmetries: } Topological phases with real Bloch states appear in spinless models with space-time inversion ${\mathcal{I T}}$ ($\mathcal{T}^2=1$). The best known examples in this class are given by the 2D Euler insulators \cite{Ahn2,Unal,Zhao4,Bouhon2,Zhao5,Ezawa}, 3D real Dirac semimetals \cite{Furusaki,Zhao3} and $\mathbb{Z}_2$ nodal-line semimetals \cite{Fu,Ahn,Wieder3,Palumbo,Tiwari} where $SO(N)$ fiber bundles emerge in momentum space due to the ${\mathcal{I T}}$ symmetry. Here, Euler, Pontryagin and Stiefel-Whitney invariants replace Chern numbers\cite{Nakahara}. We now consider a novel class of models with real Bloch states that also supports chiral symmetry. In this case, $SO(N)$ real bundle gerbes \cite{Mathai,Szabo} that generalize the $SO(N)$ fiber bundles become relevant in the description as we show below.
 When a quantum system has time-reversal $\mathcal{T}$ ($\mathcal{T}^2=1$), inversion $\mathcal{I}$ and chiral symmetry $\mathcal{S}$
 the momentum-space Hamiltonian density $H({\bf{k}})$ satisfies
 \begin{eqnarray}\label{}
\mathcal{T} H({\bf{k}}) \mathcal{T}^{-1} = H(-{\bf{k}}), \hspace{0.3cm} \mathcal{I} H({\bf{k}}) \mathcal{I}^{-1} = H(-{\bf{k}}), \hspace{0.1cm}\nonumber \\
\mathcal{S} H({\bf{k}}) \mathcal{S}^{-1} =- H({\bf{k}}). \hspace{1.5cm}
 \end{eqnarray}
 The corresponding topological invariants play a role in $(2n+1)$-D gapped and $(2n+2)$-D gapless phases with $n\geq 1$. Their existence relies on the real representation of the Clifford algebra in terms of $2^{n+2} \times 2^{n+2}$ Dirac matrices.
 We now provide two explicit examples for $n=1$.
 In 3D, the above symmetries are supported, for instance, by the following model
 \begin{eqnarray}\label{3DDirac2}
 H_{3D}= \gamma_1 \sin k_x + \gamma_2 \sin k_y+\gamma_3 \sin k_z+\\ \nonumber 
 \gamma_4 (m-\cos k_x-\cos k_y-\cos k_z),
 \end{eqnarray}
 $\gamma_1=\sigma_x \otimes \sigma_x \otimes \sigma_x$, $\gamma_2=\sigma_x \otimes \sigma_x \otimes \sigma_z$, $\gamma_3=\sigma_x \otimes \sigma_z \otimes \sigma_0$ and $\gamma_4=\sigma_y \otimes \sigma_z \otimes \sigma_y$. These $8 \times 8$ Dirac matrices are related to the real representation of the Clifford algebra. As a consequence, the system is characterized by real Bloch wavefunctions. However, differently from the 2D Euler insulators, in this case $SO(2)$ is replaced by $SO(4)$ and we do not have any well-defined Euler invariant to describe the 3D bulk. 
 By employing our new theoretical framework, we can now construct an $SO(4)$ ${\bf{C}}_{ijk}$ connection, which is naturally associated to a non-Abelian real bundle gerbe in the first Brillouin zone.
For the four-fold degenerate lower band $E_{-}$,
the corresponding Higgs field can be built from the matrix $G=\sigma_0 \otimes \sigma_0 \otimes \sigma_2$ (like in the previous cases, also this matrix is related to the 2D inversion in $k_x$ and $k_y$) such that the higher-tensor Berry connection is given by
\begin{eqnarray}\label{}
{\bf{C}}_{xyz}=\frac{8}{|E|^4} (\cos k_y \cos k_z + \cos k_x \cos k_y+\cos k_x \cos k_z \nonumber \\
- m \cos k_x \cos k_y
\cos k_z). \hspace{2.5cm}
\end{eqnarray}
This tensor field allows us to derive the topological invariant for the 3D gapped bulk through the generalized Berry-Zak phases
\begin{eqnarray}\label{}
\Upsilon_{\bf{C}}(\mathbb{T}^3)= -8,  \hspace{0.3cm} |m|<1, \hspace{0.3cm} \nonumber \\
\Upsilon_{\bf{C}}(\mathbb{T}^3)= 4,  \hspace{0.3cm} 1<|m|<3,
\end{eqnarray}
with $\Upsilon_{\bf{C}}(\mathbb{T}^3)=0$ for $|m|>3$.
As shown in Fig.(\ref{fig2}), ${\bf{C}}_{xyz}$ has maximum at the $\bf{K}$ points where the gap closes, i.e. at $\Gamma$ ($m=1$) and $X$ ($m=3$) points, respectively.
Since the $\mathcal{IT}$ symmetry in 3D  gapped phases is not related to any 3D strong topological invariant \cite{Zhao2}, then we deduce that $\Upsilon_{\bf{C}}(\mathbb{T}^3)$ is related to the presence of the chiral symmetry. 
This model is then different from 3D weak Stiefel-Whitney insulators \cite{Ahn}, which do not need the $\mathcal{S}$ symmetry and can be seen as stacks of 2D Euler insulators.
In analogy to conventional 3D chiral-invariant topological insulators with even topological invariant \cite{Hosur} and nonsymmorphic Dirac insulators \cite{Wieder}, our model supports gapless boundary states given by doubly degenerate (real) Dirac cones. In a slab geometry, one can open a boundary gap by introducing boundary terms that break $\mathcal{S}$ but preserve $\mathcal{C}_2 \times \mathcal{T}$ (here, $\mathcal{C}_2$ is the inversion symmetry on the 2D boundary and is also supported by the 3D bulk). In this way we obtain an \emph{half} Euler insulator with $e_1=1$. The existence of this gapped boundary is one of the main features of the above 3D topological phase.\\
We now define a gapless $\mathcal{IT}$- and $\mathcal{S}$-symmetric system in 4D. The existence of $\gamma_5= \sigma_z \otimes \sigma_0 \otimes \sigma_0$ allows us to introduce a Dirac model for 4D Dirac semimetals in the real representation. Formally, the linearized momentum-space Hamiltonian is similar to that in Eq.(\ref{4DDirac}) and the Higgs field is built by employing the same matrix $G$ considered in the previous 3D case. A straightforward calculation yields
\begin{eqnarray}\label{}
\Upsilon_{\bf{C}}(\mathbb{S}^3) = -4,
\end{eqnarray}
such that its absolute value provides the topological charge of the \emph{4D real Dirac cone}. Also in this case, the chiral symmetry protects the stability of this topological number similarly to the case of 4D tensor monopoles\cite{Palumbo2,Palumbo3,Palumbo4}. In fact, this four-dimensional phase can be seen as a stack of the $\mathcal{IT}$- and $\mathcal{S}$-symmetric 3D gapped phases.
Although 4D systems cannot appear in real solid-state systems, this higher-dimensional model could be realized in artificial systems such as topoelectric circuits \cite{Price}.\\

\noindent {\bf Conclusions and outlook: } Summarizing, in this work we have presented a generalization of non-Abelian Berry connections built from momentum-space Higgs fields that allow us to define higher-dimensional versions of the non-Abelian Berry-Zak phases. Through these new fields we have shown that the topological invariants of several 2D, 3D and 4D models such as QSH insulators, Euler insulators, 3D and 4D Dirac semimetals can be computed within an unified framework. Moreover, our new theoretical concepts do not only unveil the presence of generalized gauge-theory structures in band theory but also the existence of a new class of models characterized by $\mathcal{IT}$ (with $\mathcal{T}^2=1$) and $\mathcal{S}$ symmetries. We have provided two explicit models in 3D and 4D although their generalization with higher number of bands and in higher dimensions is possible. Several directions will be considered in future work. In particular, we will extend our formalism to higher-dimensional non-Hermitian topological systems, nodal-line and nodal-surface semimetals, higher-spin fermion models \cite{Bradlyn,Goldman2,Hu,Hu2} and higher-order topological phases \cite{Palumbo6}. Moreover, via a many-body generalization of tensor gauge connections we should be able to analyze topological phases of 3D interacting models within our framework.

\noindent {\bf Acknowledgments: }
The author is pleased to acknowledge discussions with Nathan Goldman, Benjamin J. Wieder, Barry Bradlyn and Gregory A. Fiete.\\

\noindent Note: After this paper has been completed, I have been informed of a recent Ref.\cite{Zhao} that discusses a gapped 3D model similar to that one in Eq.(\ref{3DDirac2}).

\appendix

\bibliography{References}






\end{document}